\documentclass[%
reprint,
amsmath,amssymb,
aps,
]{revtex4-2}

\usepackage{xcolor}
\usepackage{graphicx}
\usepackage{dcolumn}
\usepackage{bm}
\usepackage{hyperref}

\usepackage{amsmath}
\usepackage{float}
\usepackage{amsfonts}
\usepackage{soul}

\hypersetup{
	colorlinks=true,
	linkcolor=blue,
	filecolor=black,
	anchorcolor=black,
	citecolor=blue,
	filecolor=black,
	urlcolor=blue
}

\newcommand{\im}{{\rm i}}

\begin{document}
	
	\preprint{APS/123-QED}
	
	\title{
		4D topological textures in light} \author{David Marco\textsuperscript{1, 2}}
	\author{Miguel A. Alonso\textsuperscript{1, 3, 4, 5,}}
	\email{miguel.alonso@fresnel.fr}
	\affiliation{%
		\textsuperscript{1}Aix Marseille Univ., CNRS, Centrale Marseille, Institut Fresnel, UMR 7249, 13397 Marseille Cedex 20, France\\
		\textsuperscript{2}Instituto de Bioingeniería, Universidad Miguel Hernández de Elche, 03202 Elche, Spain\\
		\textsuperscript{3}The Institute of Optics, University of Rochester, Rochester, NY14627, USA\\
		\textsuperscript{4}Center for Coherence and Quantum Optics, University of Rochester, Rochester, NY14627, USA\\
		\textsuperscript{5}Laboratory for Laser Energetics, University of Rochester, Rochester, NY14627, USA
	}%
	
	\begin{abstract}
		We present 4D topological textures in (quasi)monochromatic nonparaxial optical lattices that contain all possible polarization ellipses with every combination of ellipticity and orientation in 3D space. These fields span the nonparaxial polarization space (a complex projective plane) and a 4-sphere within specific spatiotemporal regions, forming 4D skyrmionic structures. Constructed from five plane waves with adiabatically varying relative amplitudes, they are experimentally realizable in free space by focusing a temporally variant beam with a high numerical aperture lens.
	
	\end{abstract}
	
	\maketitle

\begin{figure*}
		\centering
		\includegraphics[width=0.98\textwidth]{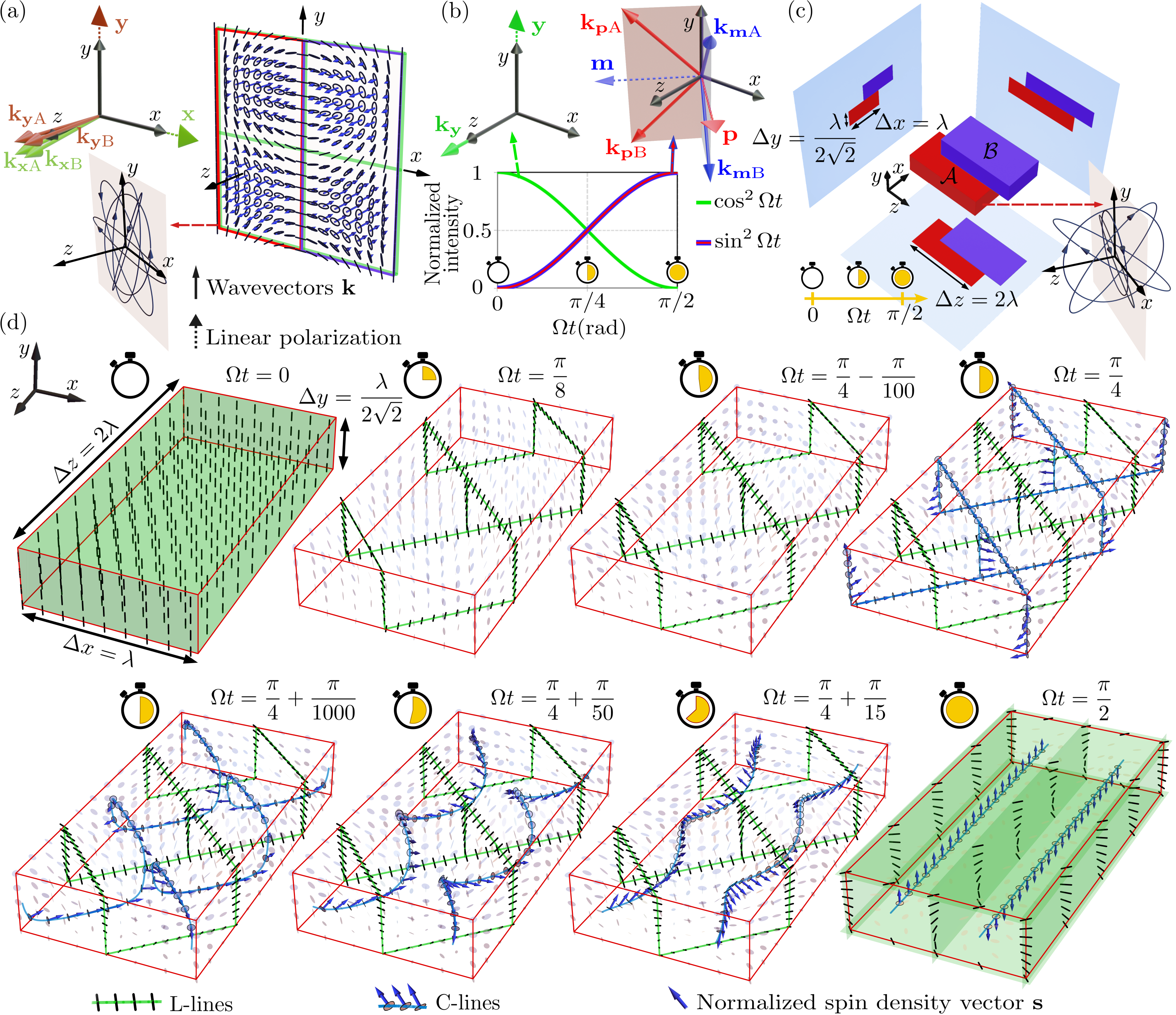}
		\caption{(a) Paraxial skyrmionic lattices formed by superimposing a small number of nearly parallel plane waves. These lattices contain regions (red and violet) where all possible polarization ellipses constrained to a plane are present, spanning the 2D paraxial polarization space (the Poincaré sphere). (b-d) In the nonparaxial regime, analogous lattices exhibit spatiotemporal polarization variations, cotaining all possible polarization ellipses with any combination of ellipticity and orientation in 3D space, within a spatiotemporal region (Full 3D Polarization (F3DP) cell), spanning the 4D nonparaxial polarization space. (b) The field is generated by adiabatically varying the relative amplitudes of two plane wave sets (green, and red and blue) at frequency $\Omega$. (c) The unit cell contains two F3DP cells ($\mathcal{A}$ and $\mathcal{B}$), each spanning differently the space of nonparaxial polarization within the temporal interval $\Omega t \in [0, \pi/2]$. (d) Temporal evolution of the polarization ellipses in F3DP cell $\mathcal{A}$.}
		\label{fig:constructingthefield}
	\end{figure*}

	\textit{Introduction}.\,---
	Skyrmionic textures are topological structures where an abstract spherical parameter space is mapped completely and monotonically onto a physical region \cite{Skyrme, cartography_skyrmions}. While magnetic skyrmions are of interest for potential data storage applications \cite{magnetic_skyrmions_review}, such textures were recently found in systems like Bose-Einstein condensates \cite{hansen2016singular} and classical wave fields \cite{review_optical_skyrmions_Shen,Ge_Acoustic, Muelas, water_skyrmions}. 
	In monochromatic optical fields, the electric field vector traces an ellipse over time, with its ellipticity and orientation defining the field’s polarization \cite{Born_Wolf}. For paraxial fields, the ellipse is essentially restricted to the plane perpendicular to the main propagation direction, so that polarization is fully described by two parameters: the ellipticity and the orientation angle within the plane. These quantities are well represented by the two coordinates over the surface of a unit 2-sphere known as the Poincaré sphere. Full Poincaré beams \cite{full_Poincare_beams}, which contain all possible paraxial polarization states in each transverse plane, map the entire Poincaré sphere onto the plane via stereographic projection, making them 
	2D optical skyrmions \cite{Poincare_skyrmions}. Superpositions of few nearly-parallel plane waves with different polarizations produce paraxial skyrmionic lattices, 
	where regions in each transverse plane map onto the Poincaré sphere \cite{prop_invariant_meron_lattices} (Fig.~\ref{fig:constructingthefield}(a)). Higher-dimensional topological structures such as Hopfions are well established in magnetic materials \cite{beyond_skyrmions} and were recently identified in optical fields that span a 3-sphere in  polarization and phase  throughout the propagation volume \cite{Dennis_hopfion}.

	For nonparaxial optical fields 
	all three Cartesian components may take on non-negligible values, so the polarization ellipse can have any 3D orientation at different points \cite{tutorial_Alonso}. Describing this elliptic shape requires four parameters, such as the ellipticity and the three Euler angles that determine its orientation. Thus, the space of nonparaxial polarization is four-dimensional. 
	Various topological structures have been found in different aspects of the polarization ellipse within spatially variant nonparaxial polarization distributions, including Möbius strips \cite{mobius_freund,mobius_polarization_experiment}, knots \cite{knots_polarization} and 2D skyrmions \cite{first_optical_skyrmions,skyrmion_spin_evanescent,skyrmions_Rodrigo_Pisanty,optical__merons_PRL}. While these structures span subspaces of the 4D space, topological textures that fully cover the 4D nonparaxial polarization space remain, to our knowledge, unexplored.

		We consider here these 4D topological structures, formed by fields spanning the space of nonparaxial polarization. This space corresponds to a manifold known as the \textit{complex projective plane}. We start by introducing a number to characterize coverage of this space.
		We also consider a 4D Skyrme number to describe 4D skyrmionic distributions, i.e., topological structures arising in fields that span a 4-sphere within a spatiotemporal region. We then find 
		(quasi)monochromatic nonparaxial optical fields 
		that fully cover both these 4D spaces, presenting 
		all possible nonparaxial polarization ellipses 
		(hence being nonparaxial counterparts of paraxial polarization lattices spanning the Poincaré sphere \cite{prop_invariant_meron_lattices,cartography_skyrmions}, as shown in Fig.~\ref{fig:constructingthefield}(a)). 
		The proposed fields are optical lattices with a polarization state distribution that changes periodically in 3D space and also (adiabatically) in time, ensuring that 
		the complex projective plane 
		is entirely covered within a cuboid in 3D space during an interval equal to a quarter of the polarization lattice’s temporal period,
		and additionally that a 4-sphere is fully covered within the same volume over twice this interval, 
		thus also producing 4D skyrmionic textures.
		These structures are generated by slowly modulating in time the relative amplitudes of a specific set of five non-coplanar plane waves, producing fields that exhibit a rich and complex subwavelength polarization structure while also having uniform intensity in space and time. Experimental implementations of these fields should be easily realizable in the focal region of a high NA microscope objective. 

	\textit{4D topological textures}.\,---
	The magnitude of the Skyrme number for a 2D texture indicates how many times the spherical parameter space is covered within a 2D region, while its sign denotes the orientation of the covering. For paraxial optical textures spanning the Poincaré sphere, it is defined as $N_\textrm{S} = (4\pi)^{-1} \iint \rho_\textrm{S} (x,y) \; \mathrm{d}x \, \mathrm{d}y$, where \(\rho_\textrm{S} (x,y) = \mathbf{n}(x,y) \cdot \left[\partial_x \mathbf{n}(x,y) \times \partial_y \mathbf{n}(x,y) \right]\) is the Skyrme density, and $\mathbf{n}=(n_1,n_2,n_3)$ is the normalized Stokes vector defining a point on the Poincaré sphere. In the unit cell of the paraxial field shown in Fig.~\ref{fig:constructingthefield}(a), the sphere is spanned twice in opposite senses, yielding $N_\mathrm{S}=\mp1$ in the red and violet regions, respetively. 
	We now 
	explore analogous phenomena for 4D textures in the nonparaxial regime.
	
\begin{figure}
		\centering
		\includegraphics[width=0.48\textwidth]{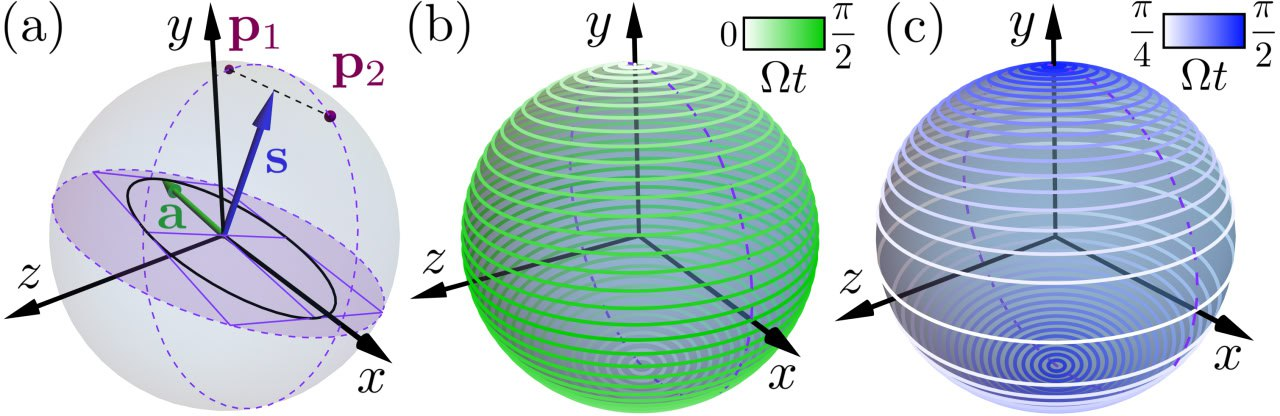}
		\caption{(a) Poincarana  construction for the polarization ellipse. The 3D orientation and ellipticity of a polarization ellipse are described by two indistinguishable points $\mathbf{p}_{1,2}$ on the unit sphere. The normalized spin density vector $\mathbf{s}$ is the centroid of the two points, and the major semi-axis vector $\mathbf{a}$ is parallel to their separation. 
			Temporal evolution on the unit sphere in real space of (b) the electric field vector direction for linearly polarized states, and (c) the normalized spin density vector for circularly polarized states.     
		}
		\label{fig:direction_sphere}
	\end{figure}

	A 3D polarization ellipse can be fully characterized by the positions of two indistinguishable points, $\mathbf{p}_1$ and $\mathbf{p}_2$, on the unit sphere ($|\mathbf{p}_1|=|\mathbf{p}_2|=1$). Different variants of this two-point construction exist \cite{Majorana,Poincarana,tutorial_Alonso}. 	
	For all, the bisector of $\mathbf{p}_1$ and $\mathbf{p}_2$ is parallel to the normal to the ellipse's plane, the line joining $\mathbf{p}_1$ and $\mathbf{p}_2$ is parallel to the ellipse's major axis, and the angular separation between $\mathbf{p}_1$ and $\mathbf{p}_2$ is a monotonic function of the ellipticity: the two points coincide for circular polarization, and they are antipodal for linear polarization. 
	Here, we adopt the so-called 
	\textit{Poincarana} representation \cite{Poincarana} (Fig.~\ref{fig:direction_sphere}(a)), which is naturally linked to geometric phase for nonparaxial fields and simplifies our calculations. In this construction, the midpoint between $\mathbf{p}_1$ and $\mathbf{p}_2$ matches the normalized spin density vector \(\mathbf{s} = \mathrm{Im}\:(\mathbf{E}^* \times \mathbf{E})/|\mathbf{E}|^2\), which is normal to the plane of the ellipse and whose magnitude is proportional to the area of the normalized ellipse, ranging from 0 (linear polarization) to 1 (circular polarization). The Poincarana points are given by $\mathbf{p}_{1, 2} = \mathbf{s} \pm \sqrt{1 - |\mathbf{s}|^2} \mathbf{a}/|\mathbf{a}|$, where $\mathbf{a}=\mathrm{Re}\{\exp{\left[-\im \mathrm{Arg}(\mathbf{E}\cdot\mathbf{E})/2\right]}\mathbf{E}/|\mathbf{E}|\}$ is the ellipse's major semi-axis   \cite{tutorial_Alonso}.
	
	For a field to span all 3D polarization states, $\mathbf{p}_1$ and $\mathbf{p}_2$ 
	must span the sphere independently, covering all combinations of the two point positions. This leads us to define a topological number for the 4D space as
	\begin{equation}
		N  =\frac{1}{2(4\pi)^2}  \iiiint \rho \, \mathrm{d}x \, \mathrm{d}y \, \mathrm{d}z \, \mathrm{d}t,
		\label{eq:SkyrmeNumber4D}
	\end{equation}
	where
	\begin{equation}
		\rho= 
		\epsilon_{i j k l}  \left[ \mathbf{p}_1 \cdot \left(\partial_i \mathbf{p}_1 \times \partial_j \mathbf{p}_1 \right)\right] \left[\mathbf{p}_2 \cdot \left(\partial_k \mathbf{p}_2 \times \partial_l \mathbf{p}_2 \right)\right]
		\label{eq:SkyrmeDensity4D}
	\end{equation}
	is the Jacobian determinant between the coordinates in the abstract
	space (the two points over the sphere) and 
	physical 
	space-time, 
	and it plays the role of the Skyrme density for the space of 3D polarization. Here, we use Einstein's convention of implicit sum over repeated indices, with $\epsilon_{ijkl}$ being the Levi-Civita tensor (which differs from zero only for the 24 terms out of 64 for which all indices are different), and $\partial_i$ for $i=1,2,3,4$ represents a derivative in $x,y,z,t$, respectively. 
	The factors of $1/2$ and $1/(4\pi)^2$
	account, respectively, for the points' indistinguishability and for the area of the sphere that each of them covers. 
	Due to the properties of the cross product, the 24 nonzero terms reduce to the 6 terms corresponding to $(i, j, k, l) \rightarrow (x, y, z, t)$, $(z, t, x, y)$, $(x, z, y, t)$, $(x, t, z, y)$, $(z, y, x, t)$, $(t, y, z, x)$, multiplied by a factor of 4. For a field that covers the space of nonparaxial polarization within a spatiotemporal region an integer number of times in the same sense, the magnitude of this integer corresponds to $|N|$, while the sense of wrapping determines the sign of $N$. This number is valid for any representation of a nonparaxial polarization ellipse in terms of two indistinguishable points on the unit sphere \cite{Majorana, Poincarana, tutorial_Alonso}.	
	
	As mentioned earlier, the space of nonparaxial polarization 
	is a 2D complex (4D real) manifold called the complex projective plane, defined by identifying triplets of complex numbers (vector $\mathbf{E}$) that differ only by an overall complex factor (the field's global amplitude and phase) \cite{roadmap_structured_waves}. In contrast, we now consider a hypothetical 4D skyrmionic texture, defined as a field that spans a 4-sphere 
	within a spatiotemporal region. The Cartesian coordinates of the 4-sphere  embedded in a 5D ambient space are the components of the vector $\mathbf{n}=(n_1,n_2,n_3,n_4,n_5)$. This texture is associated with a 4D Skyrme number, $N_\mathrm{S}$, which is an integer representing the number of times the 4-sphere is covered in the same sense,  its sign indicating the sense of this covering:
	\begin{equation}
		N_\mathrm{S}  =  \frac{3}{8 \pi^2} \iiiint \rho_\mathrm{S} \, \mathrm{d}x \, \mathrm{d}y \, \mathrm{d}z \, \mathrm{d}t,
		\label{eq:SkyrmeNumber4D_fieldsphere}
	\end{equation}
	where the normalization factor $3/(8 \pi^2)$ is the inverse of the area of the 4-sphere, and where
	\begin{equation}
		\rho_\mathrm{S} = 
		\epsilon_{i j k l m} \, n_i  \, \partial_x n_j \,\partial_y n_k \,\partial_z n_l \,\partial_t n_m
		\label{eq:SkyrmeDensity4D_fieldsphere}
	\end{equation}
	is the 4D Skyrme density (the Jacobian of the mapping).  
	
	\textit{The optical field}.\,---	
	%
	At any point, the complex electric field vector \(\mathbf{E}\) of a nonparaxial (quasi)monochromatic field can be expressed as
	\begin{equation} 
		\mathbf{E} = 
		A e^{\im\Phi} \left[ \sin\delta \left(\mathbf{p}
		\cos\gamma e^{\im \phi}
		+
		\mathbf{m}\sin\gamma e^{\im \varphi}
		\right) +  \mathbf{y}\cos\delta
		\right],\label{eq:fieldparameters}
	\end{equation}
	%
	%
	where $\mathbf{p},\mathbf{m}=(\pm\mathbf{x}+\mathbf{z})/\sqrt2$, with $\mathbf{x}$, $\mathbf{y}$, and $\mathbf{z}$ representing linear polarization states along the $x$, $y$, and $z$ axes, respectively. The six real parameters $A,\Phi,\delta,\gamma,\phi,\varphi$ are functions of position, and even time as long as their variation is negligible over the scale of an optical cycle. The real electric field is
	$\mathrm{Re}\left[{\bf E}\exp{(-\im \omega t)}\right]$, where \(\omega\) is the carrier temporal angular frequency of oscillation. 
	Four parameters are required to describe a 3D polarization state, as the global amplitude $A$ and 
	phase $\Phi$ do not affect the shape and orientation of the ellipse traced by the electric field over a temporal period. In order for the field to span all polarization states, these four parameters must span independently a range of values, for example: \(0 \leq \delta, \gamma \leq \pi/2\), which account for all relative amplitudes between the three field components, and \(0 \leq \phi, \varphi < 2 \pi\), which account for all relative phases.
	Therefore,
	the field must map the complete abstract four-parameter space $(\delta,\gamma,\phi,\varphi)$ onto 
	physical space-time $(x,y,z,t)$ while satisfying Maxwell's equations in free space.
	
	
	A simple solution arises by assuming i) a constant global amplitude $A=1$, and ii) linear dependence of the polarization parameters on spatiotemporal coordinates.
	For example, \(\gamma = k y / \sqrt{2}\),  \(\phi = -k (x+z) / 2\), \(\varphi = k (x-z) /2\), where \(k=\omega/c=2 \pi / \lambda\) is the wavenumber with \(\lambda\) being the wavelength, and 
	\(\delta = \Omega t\), where \(\Omega\ll \omega\). By also letting $\Phi=k z$ we arrive at the normalized field
	\begin{align}\label{eq:fieldlabframe}
		\mathbf{E}  &= \left[
		e^{-\im k (x + z) /2} \sin{\Omega  t} \left(
		\mathbf{p}\cos{\frac{k y}{\sqrt{2}}}
		+
		\mathbf{m}\,e^{\im k x} \sin{\frac{k y}{\sqrt{2}}}
		\right) \right.\nonumber \\ +&\left.\mathbf{y}\cos{\Omega  t}
		\right] e^{\im k z } \\
		&= \frac{\sin{\Omega  t}}{2} \left[ \mathbf{p}\,\left(
		e^{\im \mathbf{k}_{\mathbf{p} \mathrm{A}}\cdot{\bf r}}+e^{\im \mathbf{k}_{\mathbf{p} \mathrm{B}}\cdot{\bf r}} \right)
		\nonumber\right.\\&-
		\left.\mathbf{m} \, \im \left(
		e^{\im \mathbf{k}_{\mathbf{m} \mathrm{A}}\cdot{\bf r}}-e^{\im \mathbf{k}_{\mathbf{m} \mathrm{B}}\cdot{\bf r}} \right)\right]
		+
		\mathbf{y}\cos{\Omega  t}\, 
		e^{\im \mathbf{k}_{\mathbf{y}}\cdot{\bf r}},
	\end{align}
	where in the second step
	the field is expressed as a superposition of five plane waves with wavevectors
	$ \mathbf{k}_{\mathbf{p} \mathrm{A},\mathrm{B}} = k \left(-1, \pm \sqrt{2}, 1\right)/2$, 
		$\mathbf{k}_{\mathbf{m} \mathrm{A},\mathrm{B}} = k \left(1, \pm\sqrt{2}, 1\right)/2$, 
	and $\mathbf{k}_\mathbf{y} =  k \left(0, 0, 1\right)$, each orthogonal to its wave's polarization 
	(indicated in the subindex). The $z$ direction
	, aligned with ${\bf k}_{\bf y}$,
	bisects the other four wavevectors
	and defines the main propagation direction. The waves' amplitudes vary slowly in time, adiabatically shifting the relative weight of the first four compared to the fifth (Fig.~\ref{fig:constructingthefield}(b)). The field intensity $|{\bf E}|^2$ is uniform across space and time.
	
	A spatiotemporal region with all nonparaxial polarization ellipses
	can be identified
	by considering the range that the coordinates must sweep to span all possible relative amplitudes and phases between the coefficients of \(\mathbf{p}\), \(\mathbf{m}\) and \(\mathbf{y}\). We refer to 
	these cells
	as \textit{full 3D polarization (F3DP) cells}. One such cell 
	is
	given by $x\in[-\lambda/2,\lambda/2)$, $y\in[0,\sqrt{2}\lambda/4]$, and $z\in[0,2\lambda)$, within a time interval $\Omega t \in [0,\pi/2]$.
	This cell can be expanded by adjusting the plane wave angles (see Supplemental Material, Sec.~\ref{sec:supplemental_1}).
	
	
	%
	Figure \ref{fig:constructingthefield}(c) depicts the field's unit cell, consisting of two stacked F3DP cells $\mathcal{A}$ and $\mathcal{B}$ (red and violet cuboids, respectively). Cell $\mathcal{A}$ corresponds to the F3DP cell just described, while $\mathcal{B}$ is shifted by $\lambda/2$ in $x$ and $z$, and by $4\lambda/\sqrt2$ in $y$. The polarization distributions in $\mathcal{A}$ and $\mathcal{B}$ are identical except for a mirroring in $y$, resulting from applying these spatial shifts in Eq.~\eqref{eq:fieldlabframe}. 
	The entire field is constructed by stacking these unit cells (see Supplemental Material, Sec.~\ref{sec:supplemental_2}).
	
	Figure \ref{fig:constructingthefield}(d) shows the evolution of the polarization ellipses inside $\mathcal{A}$, highlighting L-lines (lines of linear polarization) and C-lines (lines of circular polarization).
	For $\Omega t = 0$, the entire space is an L-volume with polarization $\mathbf{y}$. 
	At $\Omega t = \pi/2$, the zones of linear polarization are two sets of L-planes, 
	constant in $x$ and in $y$. 
	In between, for $\Omega t \in (0,\pi/2)$, the field exhibits L-lines whose positions remain fixed over time.
	Figure \ref{fig:direction_sphere}(b) shows that, for $\Omega t\in[0,\pi/2]$, the field orientation at the regions of linear polarization covers the complete unit sphere. 
	Each 
	pair of curves (parallels over the sphere), corresponding to antipodal points that belong to the same polarization state,
	represents all the values of the orientation of the linearly polarized states at a given instant. 
	
	Figure \ref{fig:constructingthefield}(d) shows the spin density vector 
	\(\mathbf{s}\) 
	along the C-lines 
	(where $\mathbf{E} \cdot \mathbf{E}=0$ \cite{Berry_Dennis_EdotE}). These lines appear at $t = \pi/(4\Omega)$ as interconnected straight lines, then splitting into two helices, which gradually straighten as time progresses, eventually forming parallel straight lines at $\Omega t = \pi/2$.
	At this time the $\mathbf{y}$ field component vanishes, so the spin at every point is in the $\mathbf{y}$ direction (normal to the main propagation direction), leading to ``photonic wheels'' with transverse spin \cite{Bliokh_Nori_transverseAM,photonic_wheels_nature}. We plot the $\mathbf{s}$ directions along the C-lines for a given temporal instant as a curve on the unit sphere in Fig.~\ref{fig:direction_sphere}(c), which shows that $\mathbf{s}$ covers the entire sphere. 
	Supplemental \textbf{Video 1} and \textbf{Video 2} illustrate the temporal evolution of the polarization of this field and of a more directional variant, respectively. More details on the 
	polarization structure are provided in Supplemental Material, Sec.~\ref{sec:supplemental_3}.
	
	
	\textit{4D topological textures in optical fields}.\,--- The sign of $\rho$ (Eq.~\eqref{eq:SkyrmeDensity4D}) is uniform within each cell, $\mathcal{A}$ or $\mathcal{B}$, but opposite between them, yielding $N=\mp1$ for cells $\mathcal{A}$ and $\mathcal{B}$, respectively, so that the nonparaxial polarization space is fully spanned once within each cell, but with opposite senses. 
	Furthermore, when the field covers the space of nonparaxial polarization twice in a specific manner it spans a 4-sphere, forming 4D skyrmionic structures. This sphere arises naturally from the real and imaginary parts of the electric field components when the common phase proportional to the main direction of propagation $z$ is factored out in Eq.~\eqref{eq:fieldlabframe}. One can 
	define 
	a 4-sphere embedded in a 5D ambient space with the Cartesian coordinates being the components of the vector $\mathbf{n}
	=(\mathrm{Re}\,E'_\mathbf{p},\mathrm{Im}\,E'_\mathbf{p},\mathrm{Re}\,E'_\mathbf{y},\mathrm{Re}\,E'_\mathbf{m},\mathrm{Im}\,E'_\mathbf{m})$, satisfying $||\mathbf{n}||^2=1$, where $\mathbf{E'}=e^{-\im k z}\mathbf{E}$. Antipodal points ($\mathbf{n}\rightarrow-\mathbf{n}$) describe the same polarization state with a global \(\pi\) phase difference. 
	Within a spatial F3DP cell $\mathcal{A}$ (or $\mathcal{B}$), the 4-sphere is spanned once within $\Omega t \in [0,\pi]$, whereas the space of nonparaxial polarization is spanned twice with a net value of $N=0$. Specifically, within the spatial cell $\mathcal{A}$ ($\mathcal{B}$), the space of nonparaxial polarization is swept once over the interval $\Omega t \in [0,\pi/2]$, with $N=-1 (1)$, while, over $\Omega t \in [\pi/2,\pi]$, it is also fully covered but in the opposite sense, yielding $N=1 (-1)$. 
	The sign of $\rho_\mathrm{S}$ (Eq.~\eqref{eq:SkyrmeDensity4D_fieldsphere}) is uniform within $\Omega t \in[0,\pi]$ for cells $\mathcal{A}$ and $\mathcal{B}$ (and reverses for $\Omega t \in[\pi,2\pi]$), giving $N_\mathrm{S}=\mp(\pm)1$ for cells $\mathcal{A}$ and $\mathcal{B}$, respectively, within $\Omega t \in[0,\pi]$ ($\Omega t \in[\pi,2\pi]$). Expressions for $\rho$ and $\rho_\mathrm{S}$, along with additional properties of the 4D textures, are provided in Supplemental Material, Sec.~\ref{sec:supplemental_4}.
	
	These textures can be understood as 4D meron lattices. Two-dimensional meron lattices are periodic structures formed by merons: regions spanning a hemisphere of the spherical parameter space that tessellate the plane \cite{cartography_skyrmions}. The paraxial optical lattices in Fig.~\ref{fig:constructingthefield}(a) are meron lattices \cite{prop_invariant_meron_lattices}, as the regions spanning the Poincaré sphere present two sub-regions each covering a different Poincaré hemisphere with a 2D Skyrme number of $\pm1/2$. Analogously, each spatial F3DP cell in this nonparaxial field can be divided into two equal regions 
	along the $z$ coordinate. Between $\Omega t \in [0, \pi/2]$, each region covers half of the 4D polarization space with $|N| = 1/2$, while for $\Omega t \in [0, \pi]$, each 
	spans half 
	the 4-sphere with $|N_\mathrm{S}| = 1/2$.
	
	\textit{Conclusion}.\,---
	We introduced a topological number to describe fields spanning the 4D space of nonparaxial polarization. Examples of such fields were identified through the superposition of five linearly polarized plane waves with slowly varying amplitudes over time, which also form 4D skyrmionic structures. As detailed in Supplemental Material, Sec.~\ref{sec:supplemental_5}, these fields can be generated in the focal region of a high numerical aperture lens by focusing a temporally varying paraxial field created using a spatial light modulator \cite{vector_setup}. The main challenge lies not in generating but in measuring these fields, as current techniques for measuring nonparaxial polarization distributions rely on scanning a nanoparticle to measure one point at a time \cite{lindfors2005degree,mobius_polarization_experiment}, which would require scanning over four dimensions for this field. It would be interesting to investigate whether fields consisting of a single F3DP cell occupying a theoretically infinite volume with $N=1$ could exist, in direct analogy to full Poincaré beams \cite{full_Poincare_beams}.

	\section*{Acknowledgement}
	This research received funding from the Agence Nationale de Recherche (ANR) through the project 3DPol, ANR-21-CE24-0014-01. D. M. acknowledges Ministerio de Universidades, Spain, Universidad Miguel Hernández and the European Union (Next generation EU fund) for a Margarita Salas grant
	from the program Ayudas para la Recualificación del Sistema Universitario Español, and funding from Ministerio de Ciencia e Innovación, Spain, PID2021-126509OB-C22.

	\bibliographystyle{unsrt}
	
	\bibliography{apssamp}

	\clearpage
	\section*{Supplemental Information}
	

	\subsection{Increasing the size of the F3DP cell}
	\label{sec:supplemental_1}
	In the main text, we noted that the spatial dimensions of the F3DP cell can be adjusted. Here, we present a more generalized version of the field, with F3DP cell sizes modifiable based on specific parameters. The field described in the main text represents a particular case of this more general formulation. In this Supplemental Material, we provide additional results for this generalized field, which can be specialized to correspond to the specific field discussed in the main text.
	
\begin{figure*}
		\centering\includegraphics[width=0.7\textwidth]{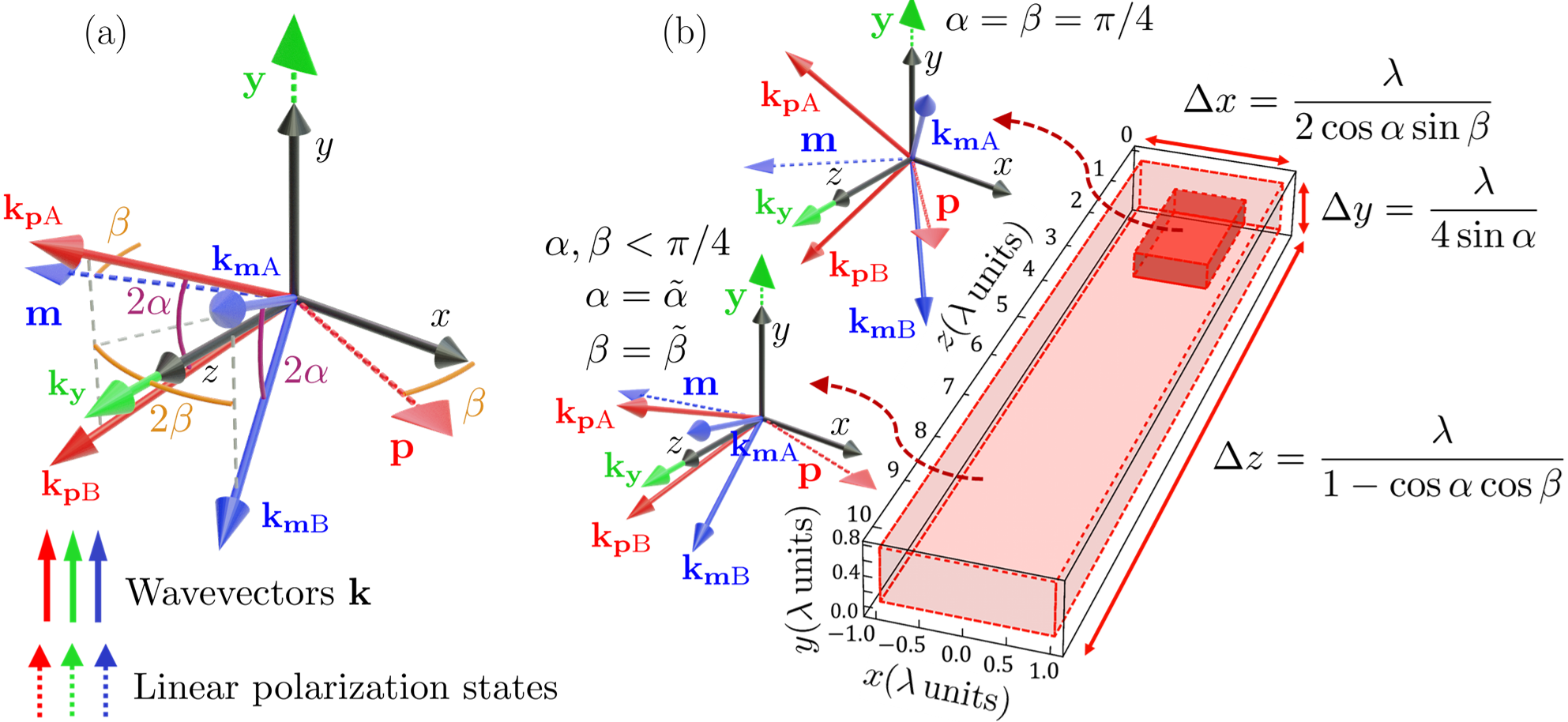}
		\caption{(a) The dimensions of the F3DP cell can be controlled by changing the angles $\alpha$ and $\beta$ for the wavevectors and the corresponding polarization vectors. The angle between two wavevectors with the same polarization state (blue-blue or red-red pairs) is \(2 \alpha\), and the angle between the projections onto the \(x z\) plane of these two pairs of wavevectors (blue-red pairs) is \(2 \beta\). (b) The size of the F3DP cell increases as \(\alpha\) and \(\beta\) decrease: the smaller cell in the figure corresponds to \(\alpha = \beta = \pi / 4\), while the larger one (whose dimensions in $x$ and $y$ are twice those of the smaller one) corresponds to \(\alpha =\tilde{\alpha}= \arcsin[1/(2\sqrt2)]=0.3614\) and \(\beta =\tilde{\beta}= \arcsin(1/\sqrt{14})=0.2706\).}
		\label{fig:increasingthesizeofthecell}
	\end{figure*}

	The F3DP cells can be enlarged by making the field more directional, that is, by bringing the wavevectors of the four plane waves with polarization states \(\mathbf{p}\) and \(\mathbf{m}\) (red and blue in Fig.~\ref{fig:constructingthefield}(b) in the main text) closer to the main propagation direction (\(z\)). It is useful to regard this transformation as a composition of the two transformations depicted in Fig.~\ref{fig:increasingthesizeofthecell}(a): First, a decrease of the angle \(2 \alpha\) subtended by each pair of wavevectors with the same polarization (blue-blue and red-red pairs in Fig.~\ref{fig:increasingthesizeofthecell}(a)). Second, a decrease in the angle 
	$\beta$ between the projection onto the \(xz\) plane of each of these four wavevectors and the $z$ axis
	(so that $2\beta$ is the angle between the projections of the blue-red pairs in Fig.~\ref{fig:increasingthesizeofthecell}(a)). The wavevectors then take the form
	\begin{subequations}
		\begin{align}
			\mathbf{k}_{\mathbf{p} \mathrm{A}} &= k \left(-\cos\alpha \sin\beta, \sin\alpha, \cos\alpha \cos\beta\right),\\
			\mathbf{k}_{\mathbf{p} \mathrm{B}} &= k \left(-\cos\alpha \sin\beta, -\sin\alpha, \cos\alpha \cos\beta\right),\\
			\mathbf{k}_{\mathbf{m} \mathrm{A}} &= k \left(\cos\alpha \sin\beta, \sin\alpha, \cos\alpha \cos\beta\right),\\
			\mathbf{k}_{\mathbf{m} \mathrm{B}} &= k \left(\cos\alpha \sin\beta, -\sin\alpha, \cos\alpha \cos\beta\right),\\
			\mathbf{k}_\mathbf{y} &=  k \left(0, 0, 1\right),
		\end{align}
		\label{eq:kvectors_alpha_beta}
	\end{subequations}
	where the particular case in the main text corresponds to \(\alpha = \beta = \pi / 4\). Note that if $\beta=\pi/4$ these wavevectors remain perpendicular to the corresponding polarization vectors (${\bf p}$, ${\bf m}$, or ${\bf y}$) regardless of the value of $\alpha$, and in fact the intensity of the field remains constant. However, changing to $\beta\neq \pi/4$ does require the rotation of the polarization vectors for the first four plane waves according to
	\begin{equation}
		\mathbf{p}, \mathbf{m} = \pm \mathbf{x} \cos{\beta} +  \mathbf{z} \sin{\beta},
	\end{equation}
	in order to ensure plane-wave transversality. Since for $\beta\neq \pi/4$ these two polarization components are not mutually orthogonal, the intensity of the field is no longer spatio-temporally uniform. The field can now be written as
	\begin{align}
		\mathbf{E}  &= \left[
		e^{-i (\kappa_z z + \kappa_x x)} \sin{\Omega  t} \left(
		\mathbf{p}\cos{\kappa_y y}
		+  \mathbf{m}\,e^{i 2 \kappa_x x} \sin{\kappa_y y}
		\right)\right. \nonumber \\
		&+ \left.\mathbf{y}\cos{\Omega  t} \right] e^{i k z} \label{eq:fieldbiggercell}\\
		&=  \mathbf{p}\,\frac{\sin{\Omega  t}}2\left(
		e^{i \mathbf{k}_{\mathbf{p} \mathrm{A}}\cdot{\bf r}}+e^{i \mathbf{k}_{\mathbf{p} \mathrm{B}}\cdot{\bf r}} \right) \nonumber \\
		&+\mathbf{m}\,\frac{\sin{\Omega  t}}{2i}\left(
		e^{i \mathbf{k}_{\mathbf{m} \mathrm{A}}\cdot{\bf r}}-e^{i \mathbf{k}_{\mathbf{m} \mathrm{B}}\cdot{\bf r}} \right)
		+
		\mathbf{y}\cos{\Omega  t}\, 
		e^{i \mathbf{k}_{\mathbf{y}}\cdot{\bf r}},
	\end{align}
	where the spatial frequencies between contributions are given by
	\begin{subequations}
		\begin{align}
			\kappa_x &= k \cos{\alpha} \sin{\beta},\\
			\kappa_y &= k \sin{\alpha},\\
			\kappa_z &= k (1-\cos{\alpha}\cos{\beta}).
		\end{align}
	\end{subequations}
	The dimensions of the F3DP cell are inversely proportional to these spatial frequencies, as shown in 
	Fig.~\ref{fig:increasingthesizeofthecell}(b), and can be made in principle arbitrarily large by making the angles $\alpha$ and $\beta$ smaller. Note that for \(\beta \neq \pi/4\) the polarization vectors \(\mathbf{p}\), \(\mathbf{m}\) and \(\mathbf{y}\) no longer define an orthogonal set, but they are still linearly independent, and all possible relative amplitudes and phases between them are swept in the cell, giving rise to all polarization states. However, for small $\beta$ some polarization states are present only in regions of relatively low intensity, and their coverage becomes less uniform.
	
	\subsection{The field's global structure}
	\label{sec:supplemental_2}
\begin{figure*}
		\centering
		\includegraphics[width=0.78\textwidth]{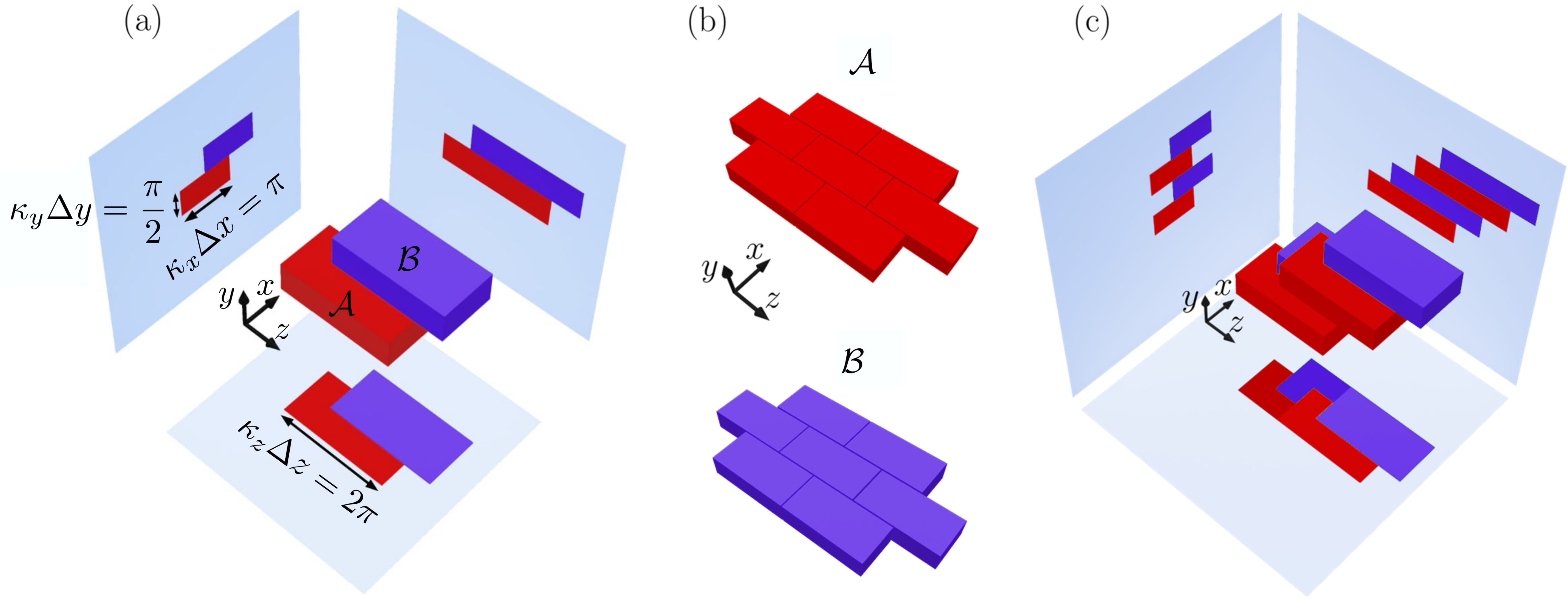}
		\caption{(a) The field unit cell consists of two F3DP cells (red and violet boxes) of dimensions $\Delta x = \pi/\kappa_x = \lambda /(2 \cos\alpha \sin\beta)$, $\Delta y = \pi/(2 \kappa_y) = \lambda /(4 \sin\alpha)$ and $\Delta z = 2 \pi/\kappa_z = \lambda /(1-\cos\alpha \cos\beta)$. The cell dimensions are expressed in units of $\kappa_j j$ ($j=x$, $y$, $z$). Projections onto the $xy$, $xz$ and $yz$ planes are shown in order to aid with 3D visualization. The (violet) cell $\mathcal{B}$ is displaced by $\pi/2$ rad in $\kappa_x x$, $\kappa_y y$ and $\kappa_z z$ with respect to the (red) cell $\mathcal{A}$, and its polarization state distribution is the same as in the red cell but mirrored in $y$. The global field structure is reproduced by stacking unit cells following two rules: stacking unit cells displaced (b) by $n\pi$ (where $n\in\mathbb{Z}$) in $\kappa_x x$ and $\kappa_z z$ (brick structure), and (c) by $n\pi$ in $\kappa_y y$ and displaced by $n\pi$ in $\kappa_x x$ or in $\kappa_z z$.}
		\label{fig:unit_cell}
	\end{figure*}

	Figure \ref{fig:unit_cell}(a) depicts a unit cell of the field's polarization distribution, which consists of two stacked F3DP cells (red and violet cuboids). The cell dimensions are expressed in units of $\kappa_j j$ ($j=x$, $y$, $z$), so the representation is valid for any value of $\alpha$ and $\beta$. The red cell represents the F3DP cell $\mathcal{A}$ described in the main text for any value of $\alpha$ and $\beta$, defined by $\kappa_x x\in[-\pi/2,\pi/2)$, $\kappa_y y\in[0,\pi/2]$ and $\kappa_z z\in[0,2\pi)$. The cell $\mathcal{B}$ (violet) is an F3DP cell displaced by $\pi/2$ in each dimensionless coordinate $\kappa_j j$ with respect to $\mathcal{A}$. It turns out that the polarization distributions in $\mathcal{A}$ and $\mathcal{B}$ are identical except for a mirroring in $y$, as one can see from applying a shift of $\pi/2$ to $\kappa_x x$, $\kappa_y y$ and $\kappa_z z$ in Eq.~\eqref{eq:fieldbiggercell}. The entire field can then be constructed by stacking these unit cells, as shown in Figs.~\ref{fig:unit_cell}(b,c). The positions where the unit cells must be stacked are deduced from the changes in $\kappa_j j$ that leave the field in Eq.~\eqref{eq:fieldbiggercell} invariant: i) a displacement of $n\pi$ ($n\in\mathbb{Z}$) in $\kappa_x x$ and $\kappa_z z$ (which results in the brick structure in Fig.~\ref{fig:unit_cell}(b) for each type of cell), and ii) a displacement of $n\pi$ in $\kappa_y y$ and a shift of $n\pi$ in $\kappa_x x$ or $\kappa_z z$ (depicted in Fig.~\ref{fig:unit_cell}(c)).
	
	\subsection{The field's polarization distribution}
	\label{sec:supplemental_3}
	
\begin{figure*}
		\centering\includegraphics[width=0.8\textwidth]{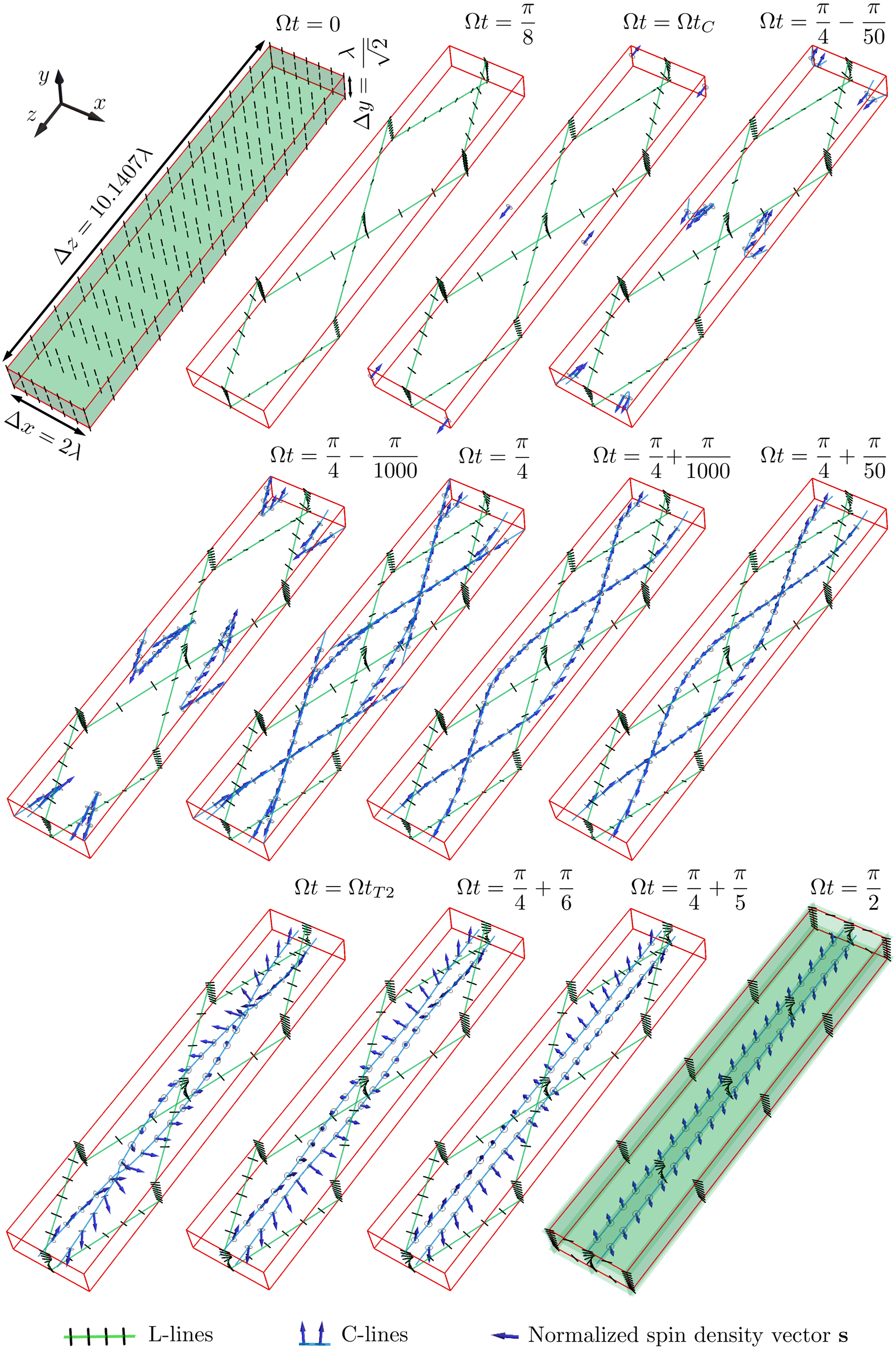}
		\caption{Temporal evolution of the L-lines and C-lines inside the F3DP cell $\mathcal{A}$ for \(\alpha = \tilde{\alpha}\) and \(\beta =\tilde{\beta}\) (for which the dimensions of the F3DP cell in $x$ and $y$ are twice as large as for the cell corresponding to $\alpha=\beta=\pi/4$). The polarization ellipses are plotted for the C-lines and L-lines, and the normalized spin density vector is plotted for the C-lines. At $\Omega t_C = 0.6331$, the C-lines appear as points. The C-lines undergo two topological transitions, occurring at $\Omega t_{T1} = \pi/4$ and at $\Omega t_{T2} =1.2094$.}
		\label{fig:CLlinesbiggercell}
	\end{figure*}

	In this section, we provide details of the derivation of the L-lines, and equations for the C-lines, the normalized spin denity vector $\mathbf{s}$, and the topological transitions that the C-lines undergo. We also study the evolution of the L-lines and C-lines for $\beta<\pi/4$, in particular for the case \(\alpha =\tilde{\alpha}= \arcsin[1/(2\sqrt2)]=0.3614\) and \(\beta =\tilde{\beta}= \arcsin(1/\sqrt{14})=0.2706\). These parameter values result in the dimensions along the $x$ and $y$ axes of the F3DP cells being twice those of the cells discussed in the main text (where $\alpha=\beta=\pi/4$). The temporal evolution of the L-lines and C-lines of the field for $\alpha=\tilde{\alpha}$ and $\beta=\tilde{\beta}$ is depicted in Fig.~\ref{fig:CLlinesbiggercell}. As we will demonstrate later,  this case exemplifies the topology of the polarization structure for all cases where $\alpha\leq\pi/4$ and $\beta<\pi/4$.
	
	\subsubsection{L-lines}
	We now describe how to obtain the L-lines for the field defined in Eq.~\eqref{eq:fieldbiggercell}. For $\Omega t = 0$, the field has a polarization state $\mathbf{y}$, which means that the 3D space itself is an L-volume. For $\Omega t = \pi/2$, on the other hand, only the components $\mathbf{p}$ and $\mathbf{m}$ survive. If, in addition, $\kappa_y y = n \pi/2$ ($n\in\mathbb{Z}$), then only one linear polarization component remains, so this condition defines planes of constant $y$ with $\mathbf{p}$ (for $\kappa_y y = n \pi$) or $\mathbf{m}$ (for $\kappa_y y = (2n+1)\pi/2$) polarization state. Conversely, for $\kappa_y y \neq n \pi/2$, the phase between the $\mathbf{p}$ and $\mathbf{m}$ components must be $n\pi$ for the state to be linear, which defines a set of L-planes $x=n \pi/(2 \kappa_x)$.
	
	Within the intermediate interval $\Omega t \in (0,\pi/2)$ we find L-lines. The first subset of L-lines is deduced for $\kappa_y y = n \pi/2$. Under this spatial constraint, only two polarization components survive ($\mathbf{y}$ and $\mathbf{p}$ or $\mathbf{y}$ and $\mathbf{m}$), and in order to have linearly polarized light, the relative phase between these components must be $n\pi$, which, together with the condition $\kappa_y y = n \pi/2$, defines a set of L-lines that lie on the planes $y=0$ and $y=\lambda/(4\sin\alpha)$ in the F3DP cell $\mathcal{A}$. The second set of L-lines is deduced for $\kappa_y y \neq n \pi/2$ (i.e., when the three linear polarization components survive), imposing that the two relative phases between the three linear polarization components must be integer multiples of $\pi$.
	
\begin{figure}
		\centering
		\includegraphics[width=0.44\textwidth]{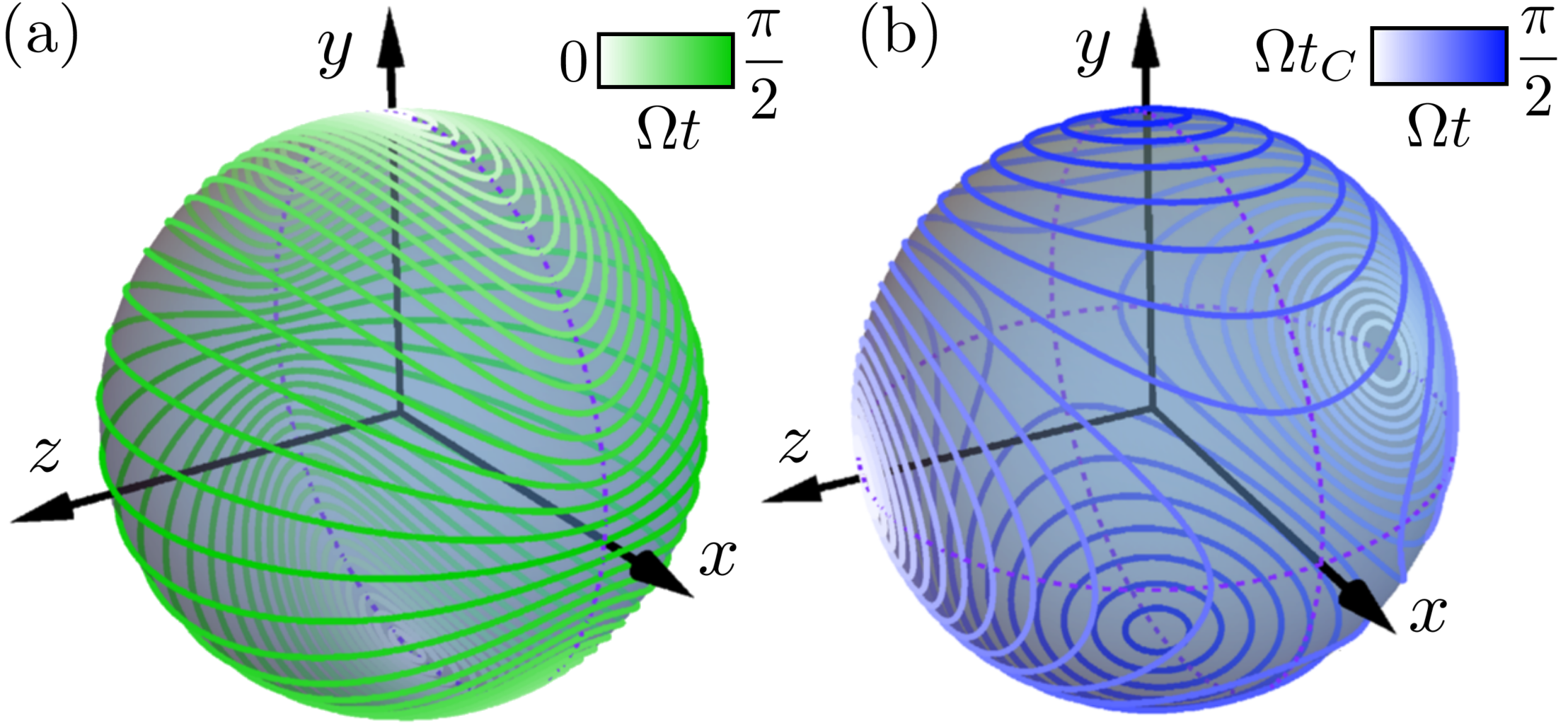}
		\caption{Temporal evolution on the unit sphere in real space of (a) the electric field vector in linearly polarized states, and (b) the normalized spin density vector in circularly polarized states for \(\alpha = \tilde{\alpha}\) and \(\beta = \tilde{\beta}\). Each curve represents a distinct moment in time, with uniformly spaced intervals between successive times. The C-lines are depicted from their initial appearance at $\Omega t_C=0.6331$.}
		\label{fig:direction_sphere_supplemental}
	\end{figure}

	For $\alpha=\tilde{\alpha}$ and $\beta=\tilde{\beta}$, the orientation of the electric field within linearly polarized regions spans the unit sphere during the interval $\Omega t\in[0,\pi/2]$, as illustrated in Fig.~\ref{fig:direction_sphere_supplemental}(a). Antipodal points represent the same linear polarization state. Each curve (or more precisely, each pair of curves) includes the orientations of linearly polarized states at a specific instant. These curves are independent of $\alpha$, and, for $\beta=\pi/4$ (Fig.~\ref{fig:direction_sphere} in the main manuscript), they are parallels. Decreasing $\beta$ (Fig.~\ref{fig:direction_sphere_supplemental}(a) for $\beta=\tilde{\beta}$) distorts the curves while maintaining full directional coverage. Over time, the $\mathbf{p}$ and $\mathbf{m}$ field components increase, while the $\mathbf{y}$ component decreases. When $\Omega t = \pi/2$, each ellipse is contained within the $xz$ plane, so the field orientation lies along the equator of the unit sphere. This explains why the curves move from the sphere's poles towards the equator as time increases.
	
	\subsubsection{C-lines and normalized spin density vector}
	
	The C-lines are found from the constraint \(\mathbf{E} \cdot \mathbf{E}=0\) \cite{Berry_Dennis_EdotE}. Since \(\mathbf{E} \cdot \mathbf{E}\) is a complex quantity, this constraint imposes two conditions, each defining a surface, and the intersections of these surfaces are the C-lines.
	
	The equations defining the C-lines for the field in Eq.~\eqref{eq:fieldbiggercell} can be found from the real and imaginary parts of $\mathbf{E}\cdot\mathbf{E}=0$:
	\begin{subequations}
		\begin{align}
			\cos^2{\Omega t} \cos{2 \kappa_z z} \nonumber \\
			+ \sin^2{\Omega t} \left( \cos{2 \kappa_x x} - \cos{2 \beta}  \sin{2 \kappa_y y} \right) = 0, \\
			\cos^2{\Omega t} \sin{2 \kappa_z z} - \sin^2{\Omega t} \cos{2 \kappa_y y} \sin{2 \kappa_x x} = 0.
		\end{align} 
	\end{subequations}
	Using \(\mathbf{s} = \mathrm{Im}\:(\mathbf{E}^* \times \mathbf{E})/|\mathbf{E}|^2\), we obtain the normalized spin density vector $\mathbf{s}$ for the field defined in Eq.~\eqref{eq:fieldbiggercell}:
	\begin{align}
		\mathbf{s} &= \frac{1}{|\mathbf{E}|^2} (S_1,S_2,S_3),\nonumber \\
		S_1&= -\sin{2 \Omega t} \sin{\beta} \left[ \cos{\kappa_y y} \sin{(\kappa_z z + \kappa_x x)}\right. \nonumber \\
		&\left. + \sin{\kappa_y y} \sin{(\kappa_z z - \kappa_x x)} \right],  \nonumber \\
		S_2&=-\sin^2{\Omega t}  \sin{2 \beta} \sin{2 \kappa_x x} \sin{2 \kappa_y y}, \nonumber \\
		S_3&= \sin{2 \Omega t} \cos{\beta} \left[ \cos{\kappa_y y} \sin{(\kappa_z z + \kappa_x x)}\right. \nonumber \\
		-&\left.\sin{\kappa_y y} \sin{(\kappa_z z - \kappa_x x)} \right],
	\end{align}
	where
	\begin{equation}
		|\mathbf{E}|^2 = 1 - \sin^2{\Omega t} \cos{2 \beta} \cos{(2 \kappa_x x ) \sin{(2 \kappa_y y )}}
	\end{equation}
	is the field intensity.
	
	For any $\alpha$ and $\beta$, there are no C-lines at $t=0$; they appear at a time $t_C$, which depends only on $\beta$, and then undergo two topological transitions within the interval $\Omega t \in [0,\pi/2]$. The first transition always occurs at the time $\Omega t_{T1} = \pi/4$, and the other one occurs at $\Omega t_{T2}$, which depends on $\beta$. 
	For \(\beta = \pi/4\) the two topological transitions coincide with the appearance of the C-lines, $\Omega t_C = \Omega t_{T1} = \Omega t_{T2}=\pi/4$, while choosing a smaller $\beta$ causes $t_C$ to decrease and $t_{T1}$ to increase. Conversely, if $\alpha$ is reduced, the C-lines get stretched out along $y$ and $z$ and shrink slightly along $x$, but $t_C$ and $t_{T2}$ remain unaltered, i.e., $\alpha$ does not alter the topological properties of the C-lines. Studying the case when $\beta = \pi/4$ and one case when $\beta < \pi/4$ is sufficient then to illustrate the topological behavior of the C-lines.
	
	For $\alpha=\tilde{\alpha}$ and $\beta=\tilde{\beta}$, the C-lines appear as points at $\Omega t_C = 0.6331$. The plot for $\Omega t = \pi/4 - \pi/50$ in Fig.~\ref{fig:CLlinesbiggercell} shows that shortly after $t_C$, the C-lines become small closed loops that spill into neighboring F3DP cells (as can be deduced from the brick-like structure in Fig.~\ref{fig:unit_cell}(b)). These loops grow until, right before $\Omega t=\Omega t_{T1}=\pi/4$, segments of them almost lie on the $y=0$ plane while others almost lie on the $y=\lambda/(4\sin\alpha)$ plane. At the topological transition at $\Omega t_{T1}=\pi/4$, pairs of C-lines segments lying on the same plane of constant $y$ get linked by straight C-lines. These straight lines come from closed C-lines loops in the upper and lower F3DP cells. Right after $\Omega t = \pi/4$, the segments almost lying on the $y=0$ and $y=\lambda/(4\sin\alpha)$ surfaces before $\Omega t = \pi/4$ disappear from the cell; they move to the F3DP cells above and below to form the same structure they form in $\mathcal{A}$ but displaced and reflected along $y$. This structure consists of two tangled helical C-lines winding around each other. At the second topological transition at $\Omega t_{T2}$, the two helical lines meet at two points, and right after $t_{T2}$, the C-line segments within $x>0$ combine to form a new C-line, and the same for $x<0$, so there are now two untangled helical C-lines. After $t_{T2}$, the C-lines behave as they did for $\beta = \pi/4$: as time increases the lines straighten gradually until they become straight for $\Omega t= \pi/2$, when the ellipses are constrained to the $x z$ plane, and pure transverse spin appears.

	For $\alpha=\beta=\pi/4$, at $\Omega t = \pi/4$, $\mathbf{s}$ spans all directions within the $xz$ plane (equator of the unit sphere in Fig.~\ref{fig:direction_sphere}(c) in the main text). After $\Omega t = \pi/4$, $\mathbf{s}$ traces identical parallels on opposite hemispheres for each of the two C-lines: at the C-line in the region $-0.5\lambda<x<0$ ($0< x<0.5\lambda$), $\mathbf{s}$ evolves in time towards the direction $+y$ ($-y$). As occurs for the curves in Fig.~\ref{fig:direction_sphere_supplemental}(a), the curves representing the $\mathbf{s}$ vector on the unit sphere do not depend on $\alpha$ but they are no longer parallels for $\beta<\pi/4$ (Fig.~\ref{fig:direction_sphere_supplemental}(b)), while still covering all possible spin directions along the C-lines within the time interval $\Omega t \in [0,\pi/2]$. Note from Fig.~\ref{fig:direction_sphere_supplemental}(b) that the topological transition at $t_{T2}$ ($\Omega t_{T2} = 1.2094$ for $\beta=\tilde{\beta}$) corresponds to topological transitions of the spin curves at $\pm{\bf x}$ over the unit sphere: the curves that cycle around the $z$ axis correspond to $t<t_{T2}$, while those that cycle around the $y$ axis correspond to $t>t_{T2}$.    
	
\begin{figure}[h!]
		\centering
		\includegraphics[width=0.33\textwidth]{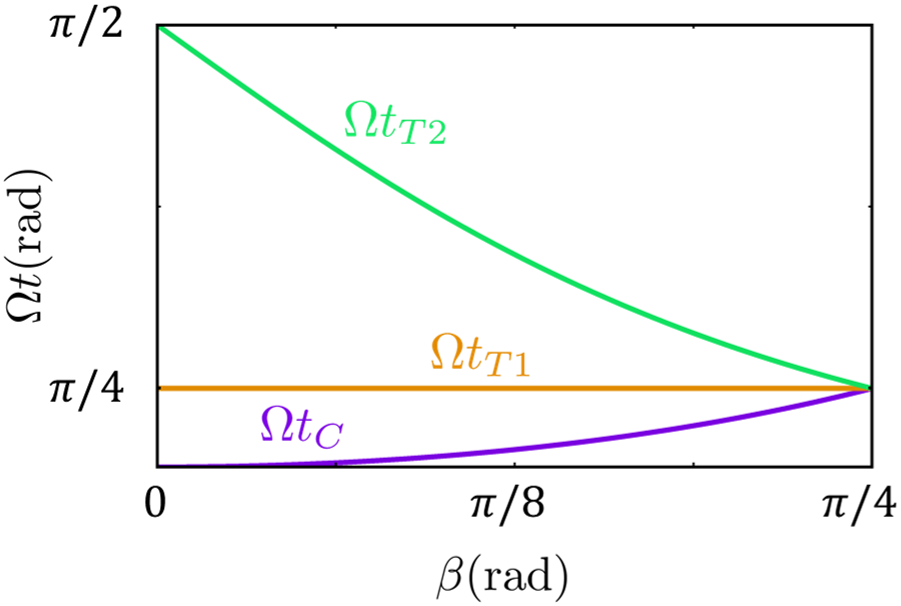}
		\caption{Dependence on $\beta$ of the time $t_{C}$ when the C-lines appear, and of the topological transition times $t_{T1}$ and $t_{T2}$.}
		\label{fig:topologicaltransitiontimes}
	\end{figure}

	The final part of this section provides further details on the topological transitions of the C-lines. For $\beta<\pi/4$, we observed for multiple values of $\beta$ that the C-lines in the F3DP cell $\mathcal{A}$ always appear as points at $x = \pi/(2\kappa_x)$, $y = \pi/(4\kappa_y)$, $z = \pi/\kappa_z$, and $x = \pi/(2\kappa_x)$, $y = \pi/(4\kappa_y)$, $z = 0$. (It is only for $\beta=\pi/4$ that they appear as lines and they also appear at other regions.) Evaluating $\mathbf{s}$ at these spatial coordinates reveals that $\mathbf{s}$ points in the $\pm z$ direction at these points, as can be seen for $\alpha = \tilde{\alpha}$ and  $\beta = \tilde{\beta}$ in Fig.~\ref{fig:CLlinesbiggercell} and on the sphere in Fig.~\ref{fig:direction_sphere_supplemental}(b). Setting $||\mathbf{s}||=1$ (which implies circular polarization) at these points, we arrive at the expression
	\begin{equation}
		\Omega t_C = \arccos\left( \frac{\cos\beta}{\sqrt{1+\frac{\cos{2\beta}}{2}}}\right)
	\end{equation}
	for the time $t_C$ when the C-lines appear: $\Omega t_C = \pi/4$ for $\beta = \pi/4$ (Fig.~\ref{fig:constructingthefield}(d) in the main text) and $\Omega t_C = 0.6331$ for $\beta = \tilde{\beta}$ (Fig.~\ref{fig:CLlinesbiggercell}). As mentioned earlier, we observed a topological transition at $\Omega t_{T1} = \pi/4 $ for any $\beta$. At a given time $t_{T2}$, there is another topological transition for the C-lines. We observed that there are two helical C-lines inside the F3DP cell $\mathcal{A}$ meeting at the points $x=0$, $y=\pi/(4\kappa_y)$ and $z=\pi/(2\kappa_z)$, and $x=0$, $y=\pi/(4\kappa_y)$ and $z= \pi/\kappa_z$ for any $\beta$. Evaluating the vector $\mathbf{s}$ in these coordinates we found that the vector $\mathbf{s}$ points along $+x$ (-$x$) for any $\beta$ at these points. Imposing that $||\mathbf{s}||=1$, we arrive at
	\begin{equation}
		\Omega t_{T2} = \arccos\left( \frac{\sin\beta}{\sqrt{1-\frac{\cos{2\beta}}{2}}}\right).
	\end{equation}
	The topological transition is noticeable from the unit sphere in Fig.~\ref{fig:direction_sphere_supplemental}(b). It occurs when $\mathbf{s}$ points along $x$ and $-x$ and it is a very fast transition. For \(\beta = \tilde{\beta}\), $\Omega t_{T2} = 1.2094$, while for \(\beta = \pi/4\), $\Omega t_{T2} = \pi/4$. As pointed before, for \(\beta = \pi/4\), the topological transitions and the appearance of the C-lines occur at the same instant $\Omega t_C = \Omega t_{T1} = \Omega t_{T2}=\pi/4$. However, as $\beta$ decreases, the instants $t_C$ and $t_{T2}$ move away from each other in time while $\Omega t_{T1}$ remains at $\pi/4$ as shown in the plot in Fig.~\ref{fig:topologicaltransitiontimes}.
	
	\subsection{Four-dimensional topological structures in nonparaxial optical fields}
	\label{sec:supplemental_4}
	
	\subsubsection{Fields spanning the 4D space of nonparaxial polarization}
	
\begin{figure}
		\centering\includegraphics[width=0.48\textwidth]{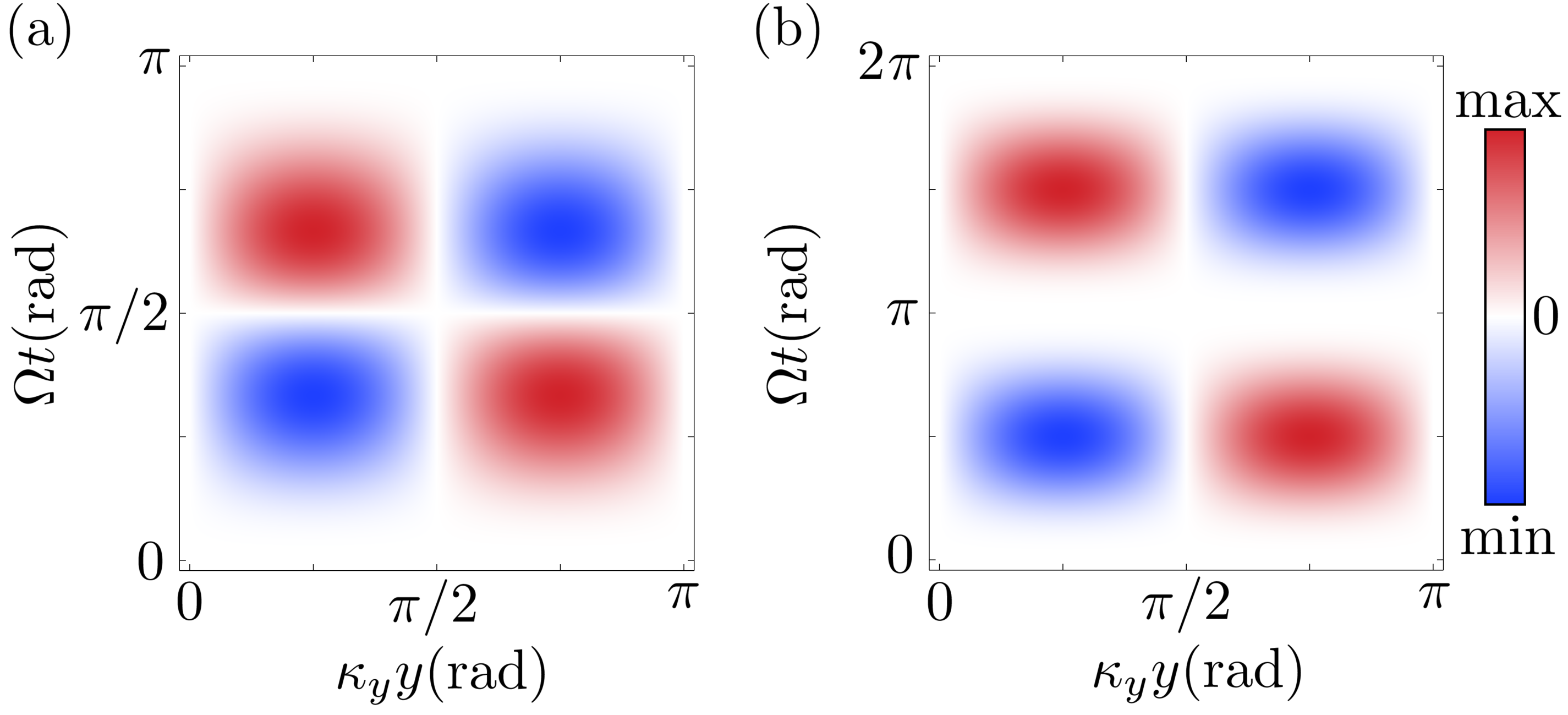}
		\caption{Unit cell of the (a) $\rho$ and (b) $\rho_\mathrm{S}$ distributions as a function of $t$ and $y$ for $\alpha=\beta=\pi/4$.}
		\label{fig:Skyrme_density}
	\end{figure}

	For any $\alpha$ and $\beta$, the sign of $\rho$, obtained from Eq.~\eqref{eq:SkyrmeDensity4D} in the main manuscript, is opposite in $\mathcal{A}$ and $\mathcal{B}$ due to the reversal of the polarization distribution in $y$. This sign also switches temporally every time $\Omega t$ is an integer multiple of $\pi/2$. We found that $\rho$ is independent of $z$ for any $\alpha$ and $\beta$, and it has periodic dependence on $x$ for \(\beta <\pi/4\) without changing sign (the sign depending only on $y$ and $t$). Further, for $\beta=\pi/4$, we observed through numeric evaluation that $\rho$ actually does not depend on $x$, and the long analytic expression for $\rho$ greatly simplifies by setting $x=z=0$, giving
	\begin{equation}
		\rho = 32 k^3 \Omega (\cos\alpha-\sqrt2)  \cos\alpha \cos{\Omega t} \sin^3{\Omega t} \sin{2 \kappa_y y}.
		\label{eq:SkyrmeDensity4DcaseI}
	\end{equation}
	This expression for $\rho$ is plotted in Fig.~\ref{fig:Skyrme_density}(a) as a function of $y$ and $t$  for $\alpha = \beta = \pi/4$. For $\beta=\pi/4$, inside a cell of type $\mathcal{A}$ or $\mathcal{B}$, $N$ can then be easily calculated analytically by integrating $\rho$ using Eq.~\eqref{eq:SkyrmeNumber4D} in the main text. This results in $N=-1$ for cell $\mathcal{A}$ and $N=+1$ for cell $\mathcal{B}$ within the interval $\Omega t \in[0,\pi/2]$. We also verified numerically that $N=-(+)1$ in cell $\mathcal{A}$($\mathcal{B}$) for several values of $\alpha$ and $\beta$. 
	Other versions of $\mathcal{A}$ and $\mathcal{B}$ displaced in $x$ and/or $z$ also have $|N|=1$, since these displacements simply shift the relative phases between the field components.
	
	Note that, since $\rho$ does not depend on $z$ for any $\beta$, $N$ accumulates linearly and indefinitely as the integration range in $z$ increases. As a consequence, one can construct cells where $N = \pm 1$ that are not F3DP cells (spatiotemporal cells containing each nonparaxial polarization state). There are also F3DP cells whose volume is the same as that for $\mathcal{A}$ or $\mathcal{B}$, but for which $N=0$. 
	
\begin{figure*}
		\centering
		\includegraphics[width=0.78\textwidth]{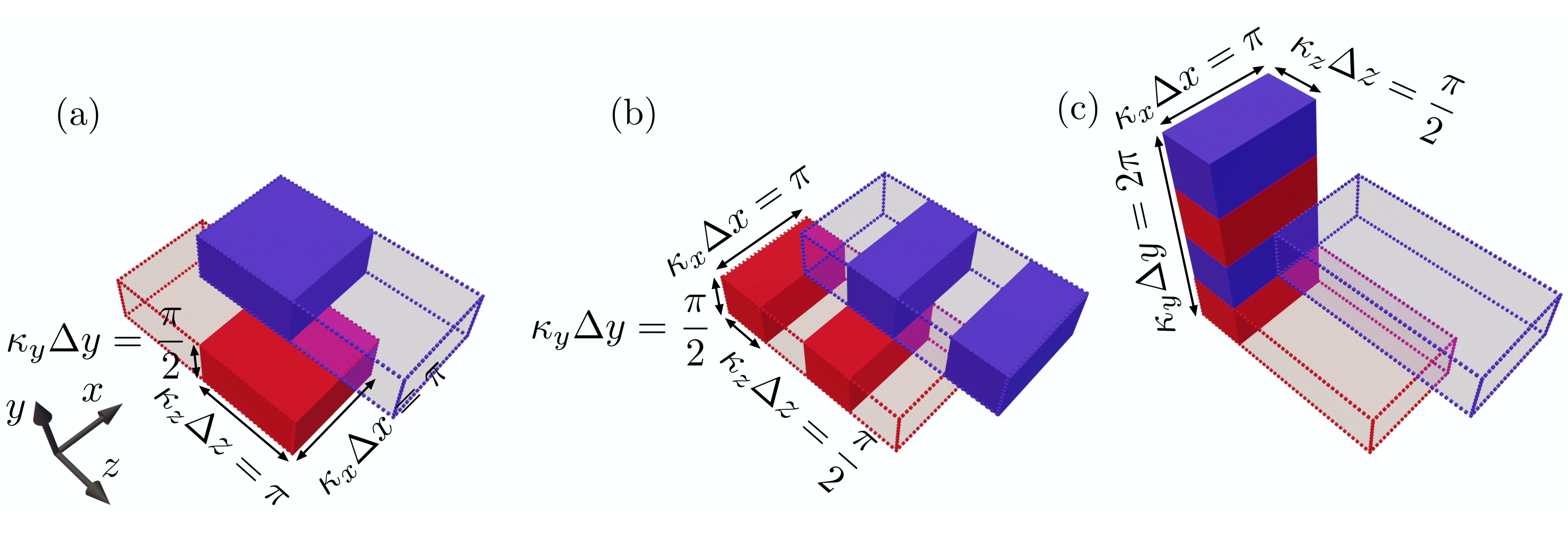}
		\caption{Regions spanning the 4D space of nonparaxial polarization within $\Omega t \in [0,\pi/2]$ but having $N=0$. These regions span the 4-sphere within $\Omega t \in [0,\pi]$ but have $N_\mathrm{S}=0$. The spatial unit cell of the polarization structure of the field has been depicted as two dashed cuboids.}
		\label{fig:nonskyrmionic_F3DP_cells}
	\end{figure*}

	
	We show in Fig.~\ref{fig:nonskyrmionic_F3DP_cells} examples of regions containing all 3D polarization ellipses where $N = 0$. The regions in (a) and (b) are constructed as follows: First, we take one or more regions in $\mathcal{A}$ that sum up to half of the $\mathcal{A}$ volume by slicing $\mathcal{A}$ along $z$. Then, we take the regions in $\mathcal{B}$ containing all the polarization states that are not in the chosen $\mathcal{A}$ regions (i.e., the regions in $\mathcal{B}$ that are not a displacement of the $\mathcal{A}$ regions by $\pi/2$ in $\kappa_x x$, $\kappa_y y$ and $\kappa_z z$). Since $\rho$ does not depend on $z$, $N$ grows linearly with $z$, and thus the total $N$ is $-1/2$ ($+1/2$) in the $\mathcal{A}$ ($\mathcal{B}$) sub-regions, so their $N$ numbers always cancel. A simple example is depicted in Fig.~\ref{fig:nonskyrmionic_F3DP_cells}(a), where we took the second half of cell $\mathcal{A}$. We can choose the $\mathcal{A}$ volumes to be split into two parts as shown in Fig.~\ref{fig:nonskyrmionic_F3DP_cells}(b). Following the stacking rules to reproduce the field (Fig.~\ref{fig:unit_cell}), one realizes that the cell in Fig.~\ref{fig:nonskyrmionic_F3DP_cells}(c) can be built from Fig.~\ref{fig:nonskyrmionic_F3DP_cells}(b), but now all the parts of the cell are connected forming a cuboidal F3DP cell with $N=0$. Analogous results can be derived for the 4D Skyrme number $N_\mathrm{S}$, given by Eq.~\eqref{eq:SkyrmeNumber4D_fieldsphere} in the main text. In the spatial regions shown in Fig.~\ref{fig:nonskyrmionic_F3DP_cells}, now constrained to the temporal interval $\Omega t \in [0,\pi]$, the 4-sphere is spanned but $N_S=0$. A similar effect is observed in the 2D paraxial texture shown in Fig.~\ref{fig:constructingthefield}(a) of the main text: in this case, the two square regions on the positive side of the $y$ axis fully cover the Poincaré sphere, yet with opposite signs of Skyrme density, resulting in a 2D Skyrme number of 0.
	
	\subsubsection{4D skyrmionic textures}
	
	The expressions defining the 4-sphere in the main text can be obtained for the general field in Eq.~\eqref{eq:fieldbiggercell}. The Cartesian coordinates of this sphere correspond to the components of the vector $\mathbf{n}=(n_1,n_2,n_3,n_4,n_5)=(\mathrm{Re}\,E'_\mathbf{p},\mathrm{Im}\,E'_\mathbf{p},\mathrm{Re}\,E'_\mathbf{y},\mathrm{Re}\,E'_\mathbf{m},\mathrm{Im}\,E'_\mathbf{m})$, where $E'=e^{-\im k z} E$ and $E$ is the field in Eq.~\eqref{eq:fieldbiggercell}. This results in the following expressions:
	\begin{subequations}
		\begin{align}
			n_1&=\sin\Omega t \cos\left(\kappa_y y\right) \cos \left(\kappa_x x + \kappa_z z\right), \\
			n_2&=-\sin\Omega t \cos\left(\kappa_y y\right) \sin \left(\kappa_x x + \kappa_z z\right), \\
			n_3&= \cos\Omega t,\\  
			n_4&=\sin\Omega t \sin\left(\kappa_y y\right) \cos \left(\kappa_x x - \kappa_z z\right), \\
			n_5&=\sin\Omega t \sin\left(\kappa_y y\right) \sin \left(\kappa_x x - \kappa_z z\right),
		\end{align}
	\end{subequations}
	with $||\mathbf{n}||^2 = 1$. Using Eq.~\eqref{eq:SkyrmeDensity4D_fieldsphere} in the main text, the 4D Skyrme density for the general field is
	\begin{equation}
		\rho_\mathrm{S} = - \kappa_x \kappa_y \kappa_z \Omega \sin{2 \kappa_y y} \sin^3{\Omega t}.
		\label{eq:SkyrmeDensity4D_4sphere}
	\end{equation}
	The Skyrme density only depends on $y$ and $t$, as illustrated in Fig.~\ref{fig:Skyrme_density}(b). Note the existence of regions in $y$ and $t$ in which the sign of the Skyrme density is flipped, this means that the sphere is being wrapped in the opposite sense. The Skyrme number inside a spatial cell of type $\mathcal{A}$ or $\mathcal{B}$ can then be easily calculated analytically by integrating the Skyrme density in Eq.~\eqref{eq:SkyrmeDensity4D_4sphere}, yielding $N_\mathrm{S}=\mp1$ for cell $\mathcal{A}$ ($\mathcal{B}$) within the interval $\Omega t \in[0,\pi]$. 
	
	Similar to what occurs for the number $N$, displacing $\mathcal{A}$ and $\mathcal{B}$ along the $x$ or $z$ directions also yield $|N_\mathrm{S}|=1$. The Skyrme number is independent of $x$ and $z$, so it increases linearly without bound as the integration limits in $x$ and $z$ extend. Therefore, cells with $N_\mathrm{S} = \pm 1$ not spanning the 4-sphere are possible.
	
	\subsection{Experimental implementation}
	\label{sec:supplemental_5}
	The field can be implemented experimentally by focusing with a microscope objective the light emerging from five points at the back focal plane, where the polarization and phase of each is appropriately prepared. The separation of the points sets the value of the angles $\alpha$ and $\beta$. The minimum numerical aperture (in air) required to produce this field is $(1-\cos^2\alpha\cos^2\beta)^{1/2}$.
	
\begin{figure}
		\centering
		\includegraphics[width=0.48\textwidth]{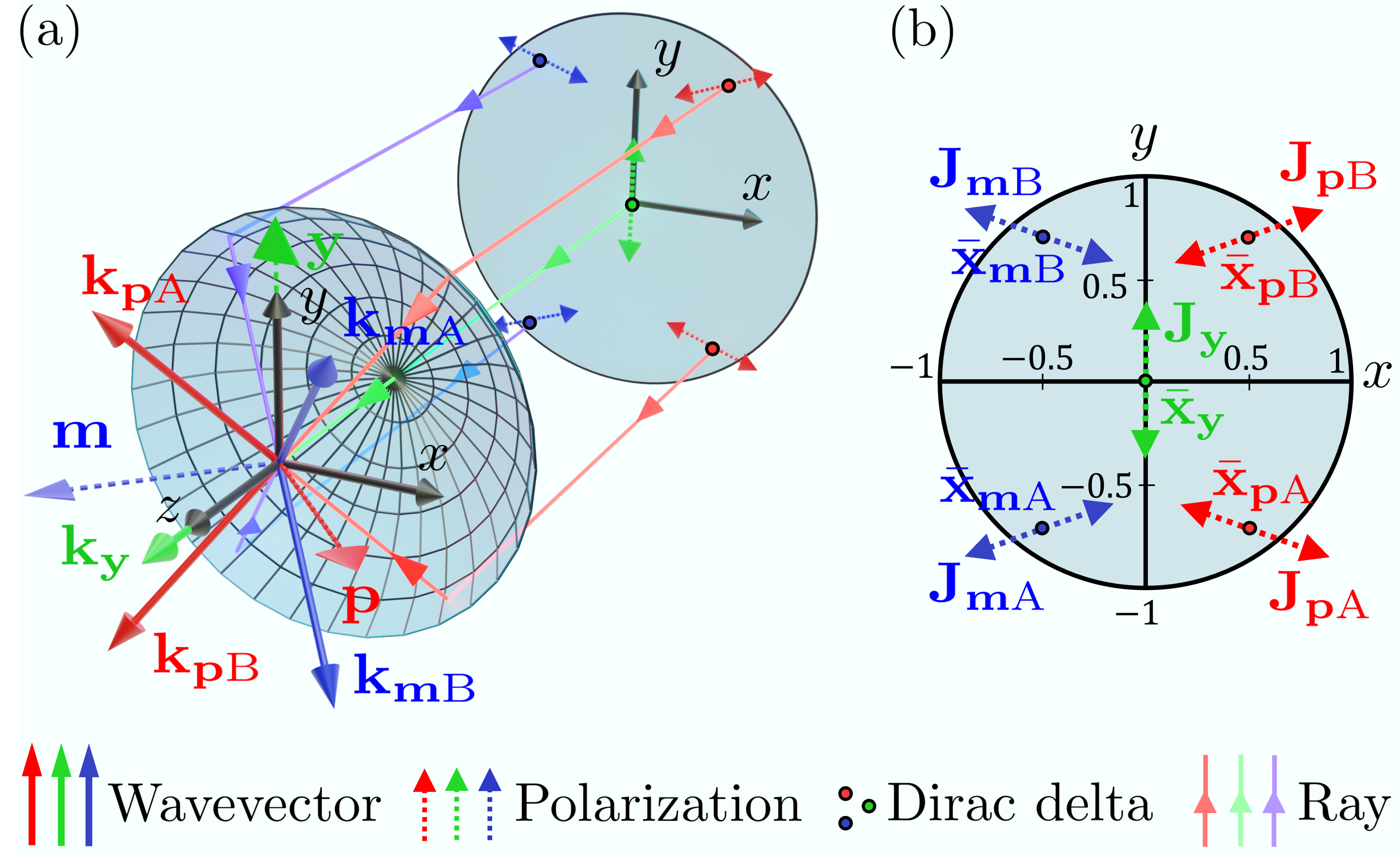}
		\caption{(a) Schematic representation of the generation of a nonparaxial field through the focalization of a paraxial field at the back focal plane of a high numerical aperture lens. (b) Illustration of the paraxial field at the back focal plane. The results are presented for the specific parameters $\alpha = \beta = \pi/4$. The generation of the field requires an adiabatic temporal amplitude variation between two sets of points as indicated in Eq.~\eqref{eq:backfocal_plane_field}.}
		\label{fig:experimental_implementation}
	\end{figure}

	The polarization state of each input point in the back focal plane is determined by rotating the polarization of the corresponding nonparaxial wave within the plane defined by its wavevector and the $z$ axis. The rotation angle is equal in magnitude but opposite in sign to the angle between these two directions, ensuring that the wavevector becomes parallel to the $z$ axis after the rotation. The positions of each point are given by the first two components of the wavevector's unit vector for each plane wave (Eqs.~\eqref{eq:kvectors_alpha_beta}) with a sign inversion. The paraxial field at the back focal plane of the microscope objective is
	\begin{align}
		\mathbf{E}=\frac{\sin\Omega t}{2}\{\delta(\mathbf{x}-\bar{\mathbf{x}}_{\mathbf{p}\mathrm{A}}) \mathbf{J}_{\mathbf{p}\mathrm{A}} + \delta(\mathbf{x}-\bar{\mathbf{x}}_{\mathbf{p}\mathrm{B}}) \mathbf{J}_{\mathbf{p}\mathrm{B}} \nonumber \\
		-\im\left[\delta(\mathbf{x}-\bar{\mathbf{x}}_{\mathbf{m}\mathrm{A}}) \mathbf{J}_{\mathbf{m}\mathrm{A}}-\delta(\mathbf{x}-\bar{\mathbf{x}}_{\mathbf{m}\mathrm{B}}) \mathbf{J}_{\mathbf{m}\mathrm{B}}\right] \} \nonumber \\
		+\cos\Omega t \,\delta(\mathbf{x}-\bar{\mathbf{x}}_\mathbf{y}) \mathbf{J}_\mathbf{y},
		\label{eq:backfocal_plane_field}
	\end{align}	
	where $\mathbf{x}=(x,y)$, and $\delta$ represents the Dirac delta distribution centered at
	\begin{subequations}
		\begin{align}
			\bar{\mathbf{x}}_{\mathbf{p}\mathrm{A}}=\left(\cos\alpha \sin\beta, -\sin\alpha\right),  \\
			\bar{\mathbf{x}}_{\mathbf{p}\mathrm{B}}=\left(\cos\alpha \sin\beta, \sin\alpha\right),  \\
			\bar{\mathbf{x}}_\mathbf{y}=\left(0,0\right), \\
			\bar{\mathbf{x}}_{\mathbf{m}\mathrm{A}}=-\bar{\mathbf{x}}_{\mathbf{p}\mathrm{B}}, \\
			\bar{\mathbf{x}}_{\mathbf{m}\mathrm{B}}=-\bar{\mathbf{x}}_{\mathbf{p}\mathrm{A}}.	\end{align}
	\end{subequations}
	The Jones vectors are
	\begin{subequations}
		\begin{align}
			\mathbf{J}_{\mathbf{p}\mathrm{A}}&=\frac{1}{f(\alpha,\beta)}\left(\cos\alpha+\cos\beta,-\sin\alpha\sin\beta\right), \\ 
			\mathbf{J}_{\mathbf{p}\mathrm{B}}&=\frac{1}{f(\alpha,\beta)}\left(\cos\alpha+\cos\beta, \sin\alpha\sin\beta\right),  \\ 
			\mathbf{J}_\mathbf{y}&=\left(0,1\right),  \\ 
			\mathbf{J}_{\mathbf{m}\mathrm{B}}&=-\mathbf{J}_{\mathbf{p}\mathrm{B}},  \\ 	
			\mathbf{J}_{\mathbf{m}\mathrm{A}}&=-\mathbf{J}_{\mathbf{p}\mathrm{A}},	
		\end{align}
	\end{subequations}
	with $f(\alpha,\beta)=1+\cos\alpha \cos\beta$.
	Figure \ref{fig:experimental_implementation} illustrates the input field for $\alpha=\beta=\pi/4$.
	
	\subsection{Videos}
	
	\label{supplementary_videos}
	\label{sec:supplemental_6}
	
	We show the evolution of the L-lines, C-lines and normalized spin density vector $\mathbf{s}$ inside the F3DP cell $\mathcal{A}$ for $\alpha = \beta = \pi/4$ (\textbf{video 1}) and for $\alpha = \tilde{\alpha}$ and $\beta = \tilde{\beta}$  (\textbf{video 2}). The evolution of the electric field orientation at the regions of linear polarization and of the $\mathbf{s}$ vector at the C-lines over the unit sphere is included.
\end{document}